\documentclass[aps,prl,twocolumn,10 pt,showpacs]{revtex4}
\usepackage{amssymb}

\usepackage{graphicx}

\begin{document}

\title{Out of equilibrium electronic transport properties of a misfit cobaltite thin film}
\author{A.Pautrat, H.W. Eng, W.Prellier}
\affiliation{CRISMAT, UMR 6508 du CNRS et de l'ENSI-Caen,6 Bd
Mar\'echal Juin, 14050 Caen, France.}

\begin{abstract}
We report on transport measurements in a thin film of the 2D
misfit Cobaltite $Ca_{3}Co_{4}O_{9}$. Dc magnetoresistance
measurements obey the modified variable range hopping law expected
for a soft Coulomb gap. When the sample is cooled down, we observe
large telegraphic-like fluctuations. At low temperature, these
slow fluctuations have non Gaussian statistics, and are stable
under a large magnetic field. These results suggest that the low
temperature state is a glassy electronic state. Resistance
relaxation and memory effects of pure magnetic origin are also
observed, but without aging phenomena. This indicates that these
magnetic effects are not glassy-like and are not directly coupled
to the electronic part.

\end{abstract}


\newpage
\maketitle
 Layered cobaltites have recently
received a growing
 interest due to their interesting properties. Their rather low resistivity
 and large Seebeck coefficient at high temperature lead to interesting thermoelectric
 properties \cite{sylvie1}. Superconductivity has been also observed in $Na_{x}CoO_{2}$, $yH_{2}O$ \cite{supra}. From a more fundamental point of view, different interesting points have been noted. For example, specific heat evidences a large effective mass in the $(Na,Ca)Co_{2}O_{4}$ family \cite{ando},
  a good indication of strongly correlated electronic properties. Strong electronic correlations were also recently reported in $Ca_{3}Co_{4}O_{9}$ using electronic transport measurements under pressure \cite{limelette}. $Ca_{3}Co_{4}O_{9}$ is a misfit incommensurate layered system ,
with small modulated atomic positions in each layer ([$CoO_2$] and
[$Ca_{2}CoO_{3}$]) \cite{riri}. Furthermore, there is a
possibility of frustrated magnetic interactions in a kagome
lattice in the $CoO_2$ layers \cite{koshi}. Ferrimagnetic or very
weak ferromagnetic-like properties are also observed at low
temperature \cite{sylvie}. In addition, $\mu^+$SR experiments have
been interpreted with the presence of a spin density wave at low
temperature \cite{Sujiyama}, giving a clue for a gap opening near
the Fermi level leading to the observed "Fermi liquid"-insulator
transition. Nevertheless, the low temperature transport behavior
is not really understood, specially because the ground state of
the system is quite complex. The eventual link between all these
observations and the measured transport properties has to be
clarified. The above mentioned results suggest that low
dimensionality, frustration and disorder, together with strong
electronic correlations, can be important features for the
physical properties of the $Ca_{3}Co_{4}O_{9}$ system. Since they
are known to reinforce fluctuations and metastability, one can
expect a strong influence on the transport properties. To our
knowledge, this has not been verified by the usual macroscopic
transport measurements. Because of the statistical averaging, the
discrete nature of the underlying "mesoscopic" processes can be
masked in a bulk sample but revealed in a small-area system. In
this case, the transport measurements can be used as an efficient
probe to bring some light on the ground states of the system
\cite{mike}.

 In this paper, we report a study of DC magnetotransport properties and of
resistance fluctuations in a microbridge of a $Ca_{3}Co_{4}O_{9}$
thin film at low temperature. We particularly focus on the origin
(magnetic or electronic) of these resistance fluctuations and
relaxation.

A $2000 \AA$ epitaxial thick film of $Ca_{3}Co_{4}O_{9}$ was used
for the measurements. The film was deposited on (0001)
$Al_{2}O_{3}$ using the pulsed laser deposition technique. The
details of the optimization, growth conditions and structural
characterizations have been described previously \cite{eng}. Prior
to the transport measurements, a silver layer was firstly
deposited via thermal evaporation onto the film, and secondly a
gold layer via RF sputtering. Microbridges ($L=200$ $\mu m$
$\times$ $W=50$ $\mu m$) were then patterned with UV
photolithography and argon ion etching. Thin aluminum contact
wires were finally used to connect the areas to the electrodes
with a wire bonding system. Measurements were made with the four
probes method. For the following results, a maximum of 100 $nA$
low noise current was used and any significant current dependence
for low current was observed, i.e., quasi-linear response
conditions were utilized. The magnetic field was applied along the
c-axis of the film, perpendicular to the substrate plane.

As it is observed in bulk samples, the $Ca_{3}Co_{4}O_{9}$ film
exhibits both a strong increase of the resistance and a large
negative magnetoresistance at low temperature \cite{masset} (we
measure $\Delta R/R \gtrsim$ -0.6 at 14 $T$ and 2 $K$). The
analysis of the transport properties in this low temperature part
is the main goal of this experiment. One of the most observable
characterization of a localized electronic state is the functional
form of the dc resistance. A simple activated law does not
describe the data, but a variable range hopping (VRH) expression
of the form $R=R_0. exp(T^{*}/T)^{\mu}$ ($\mu <$ 1) gives a very
good fit (Fig. 1) for a large temperature range (2 $K$ to 90 $K$).
Interestingly, one finds a much better agreement using $\mu =$ 1/2
than using the other exponents (1/3, 1/4) which can be expected
for the Mott VRH regime \cite{mott}. This value has been confirmed
by analyzing the so-called Zabrodski plot \cite{zabro}. This
latter consists in plotting $log(\partial R /
\partial (1/T))$ as function of $log(T)$, the slope of which gives
$\mu=$ 0.47 $\pm$ 0.02 with a robust precision and mainly without
any presupposition on the form of $R(T)$. The exponent $\mu =$ 1/2
is the most ordinary signature of a soft Coulomb gap in the
density of states (the Efros-Shklovskii or ES gap)
\cite{Efros1,Efros2}, meaning that strong Coulomb
(electron-electron) interactions are present. Theory predicts
$T^{*} = A e^2/(k_B 4\pi \varepsilon_0 \kappa \xi)$ ($k_B$ is the
Boltzmann constant, $\varepsilon_0$ is the permittivity of the
air, $\kappa$ the relative dielectric constant, $A= $2.8 and 6.2
for respectively 3D \cite{Efros2} and 2D \cite{nguyen} systems,
and $\xi$ is the localization length). This approach is a single
particle approach and neglects the correlated motion of the
charges. It was found from numerical simulations that correlations
effects on the carriers motion relies essentially on a decrease of
$A$ ($A \approx$ 0.61 was found for the 2D case in a small
system). One of the principle merit of this result is to reconcile
the introduction of many electron configurations with an unchanged
functional form of the ES-VRH \cite{perez}, but this shows that
there is a large uncertainty on $A$. From the data at $B=$ 14 $T$,
we get $T^{*}=$ 159 $K$ from $T=$ 2 $K$ to about 90 $K$. $\kappa .
\xi \approx$ 62 and 640 $nm$ for $A=$ 0.61 and 6.2 respectively
can be deduced.  $\kappa$ is to our knowledge unknown for
$Ca_{3}Co_{4}O_{9}$, but considering $\kappa \approx$ 10 as rough
approximation, reasonable $\xi$ values can be deduced. Exactly the
same analysis was applied to the zero field curve from $T=$ 90 $K$
up to $T = T_0 \approx$ 15 $K$. For $T \leq T_0$, a large negative
magnetoresistance is present. This reflects in a slight increase
of the exponent $\mu$ (0.56 $\pm$ 0.02 with the Zabroski plot).
This indicates a small departure from a pure parabolic form of the
Coulomb gap \cite{ra}. However, the most significant feature is
the change of $T^{*}$. This is this change, from $T^{*}= $159 $K$
for $B=$ 14 $T$ to $T^{*}=$ 267 $K$ for $B=$ 0 $T$, which gives
the weight to the observed negative magnetoresistance. Note that
since $\mu$ does not change significantly for our large magnetic
fields range,
 this means that the electronic interactions keep dominant with no sign of a magnetic hard gap at least for $T \gtrsim$ 2 $K$. Unlike in the 1D cobaltite $Ca_{3}Co_{2}O_{6}$
\cite{raquet}, the Coulomb gap seems to be robust to a high
magnetic field. We suggest thus that the magnetoresistance effect
comes principally from orbital, rather than spin, effect
\cite{sivan}. This agrees also with the linear field dependence of
the magnetoconductance at intermediate fields (fig.2)
\cite{nguyen2}. Nevertheless, this point deserves clearly more
attention, with a systematic study of the magnetoresistance as a
function of the magnetic field orientation. Concerning the DC
properties, it can be noted that simple activated form, together
with transport non linearities usually associated to a spin
density waves condensation and to their collective depinning
\cite{gruner} have not been observed.

We will now focus on the time series of the change in resistance.
When $T \lesssim $25 $K$, large telegraphic-like fluctuations
appear. They are generally two levels fluctuations (Fig. 3). The
square of the Fourier transform of the time traces gives their
spectral representation. After a long acquisition (several hours),
we observe a Lorentzian form, meaning that a single characteristic
time is sufficient to be statistically representative of the
switching process (Fig.4). The fluctuations are slow, a switcher
being in average about 100 $sec$ in a resistance state. These
characteristic times are only slightly temperature dependent. For
high field values (several Teslas), some sequences of three and
four levels have been also seen. At 25 $K$, the first minutes of
the resistance traces are dominated by large switches which
contribute to the noise, but after some times (typically 1000
$sec$), the noise relaxes to low value. A small perturbation as a
50 $G$ field is enough to destroy the kinetic between the two
states (Fig. 3a). Thus, the noisy regime is clearly not really
established and the rare events can be thought of precursors of
the low temperature state. In strong contrast, at 22.8 $K$, the
noisy regime is robust to a 14 $T$ field (fig.3b) and is
stabilized at least up to several hours, the typical time scale of
our longest recording. This indicates a dramatic change on the
charge dynamics. This change is confirmed by a peak in the noise
value $\sigma_R$ at a temperature $T=$ 22.9 $\pm$ 0.1 $K$
($\sigma_R$ corresponds to the integrated noise spectrum over a
10$^{-3}$-10$^{-1}$ $Hz$ bandwidth) (fig. 3c). This temperature is
reasonably separated from $T_0$, the temperature where the
magnetoresistance begins to be measurable. This is a good
indication that the noise has not a magnetic origin, in agreement
with the small sensitivity to the large magnetic fields that has
been observed. Correlated hopping conduction should be rather
involved, in agreement with the very long time scales associated
to the resistance fluctuations \cite{massey}.

  At the lowest temperatures, a notable change in the fluctuation dynamics, with the appearance of a large intermittence (Fig.4), is observed.
 After Fourier transforming and squaring the time traces, the noise
spectra are found with a $1/f^{\alpha}$ shape, where $\alpha$ is
close to 1.3 (inset of Fig.5). This contrasts to the Lorentzian
shape observed near $T=$ 22.9 $K$. We stress also that the
observed small temperature dependence of the noise power when the
temperature decreases rules out simple thermally activated
trapping processes. In addition, it can be realized that the
observed intermittence implies that the fluctuations are not
stationary anymore. This means that the noise spectrum itself
significantly changes with time. This "noise of the noise" can be
characterized by the second spectrum $S^{2}(f) = 1/f^{1-\beta}$
introduced in ref. \cite{restle}. This corresponds to the fourth
order statistic spectrum of the ordinary $1/f^{\alpha}$ noise. A
non white $S^{2}(f)$ is typical of a non Gaussian averaging
process. If one excludes the case of a dynamic redistribution of
the current in an inhomogeneous paths gallery \cite{percol}, this
implies interacting fluctuators. In practice,  the voltage signal
was acquired during a very long time (several hours) then
segmented, and finally each segment was Fourier transformed and
integrated to obtain time series of noise power. They were taken
for different ranges of frequencies and then Fourier transformed,
to finally obtain $S_2(f)$. As shown in fig.6, this second
spectrum is found nearly white at $T=$ 22.8 $K$ in the telegraphic
noise regime. In strong contrast, a large exponent $(1-\beta)
\approx$ 0.9 is measured at 8 $K$. We conclude that charges
fluctuations are strongly correlated at this moderately low
temperature. This behavior is close to what was observed in 2D
electronic systems far in the localized regime, at a much lowest
temperature \cite{bogda}. In summary, the low temperature phase of
$Ca_{3}Co_{4}O_{9}$ exhibits glassy electronic transport
properties and a Coulomb gap form of the dc conductivity, i.e.,
can be called a Coulomb Glass.

 Furthermore, when the sample is cooled down from high
temperature to 4 $K$, the resistance is measured hysteretic when
cycling the magnetic field for the first time (fig.7). It is
reversible after this first cycle and then reveals the equilibrium
magnetoresistance properties (fig.2). In addition, after a field
cooling, the resistance relaxes slowly downwards with time. A
simple exponential relaxation gives generally a good fit to the
data, and its asymptotic limit after a long time is close to the
low resistance state value. Such out of equilibrium features can
be observed in very large number of systems which exhibits pinning
by disorder, metastability, genuine glassy properties (etc). In
the case of $Ca_{3}Co_{4}O_{9}$, the possible frustration of the
interactions between spins in the triangular Co lattice
degenerates the free energy. This can be a reason for spin
glass-like properties as observed for example in Kagome lattices
\cite{ladieu}. A discriminating test is to check if the system
itself evolves with time, i.e., "ages". For that, a standard
procedure is started at a temperature $T_i$ and at a field $B_i$,
and then the system evolves with the time $t_w$. The field is
quenched and the resistance is measured during the time $t$
\cite{vincent}. After this procedure, if aging occurs, the
relaxation should differ for different $t_w$. A scaling law in
$t/t_w^{\alpha}$ provides generally a good fit to the data
\cite{vincent}. In the inset of the fig.7, it can be observed that
the system evolves during the waiting time, but with a relevant
rescaling in $t-t_w$. This means that the system relaxes but
respects time-translation invariance and that the magnetic out of
equilibrium state $\textit{is not}$ a spin-glass in its proper
sense. Nevertheless, a memory of the relaxation after a small
temperature change is seen. It is known that similar experimental
signatures, namely slow relaxation and memory without aging, are
observed when weakly interacting magnetic nanoparticles are
present, without the need of a correlation length of a spin-glass
order \cite{sasaki}. Thus, we propose an explanation of the
relaxation properties in terms of superparamagnetic-like
relaxation. The associated relaxation time $\tau^{-1}$ is maximal
at the temperature $T_0$ where the magnetoresistance has been
found to disappear (not shown), confirming the magnetic origin of
the relaxation. We propose that short range ordered clusters of
mostly ferromagnetic Co ions are the relevant entities. It is
clear that any probe of magnetic correlations at intermediate
scales, such as probed by Small Angle Neutron scattering, should
be interesting to give more clues on this magnetic ground state.

 In summary, we have shown that the low temperature magnetotransport properties of $Ca_{3}Co_{4}O_{9}$ can be
 described as a localized state with a Coulomb gap. A non
 Gaussian regime of resistance fluctuations is present at moderately low temperature. This can be attributed to
 correlated charge fluctuations characteristics of an electron glass. An additional field dependent resistance relaxation is observed
 but does not show glassy phenomena like aging, and thus is not directly coupled to the pure electronic part.

Acknowledgments: A.P would like to acknowledge  L.
M$\acute{e}$chin (GREYC-ENSICAEN) for patterning the microbridges,
S. H$\acute{e}$bert and Ch. Simon (CRISMAT-ENSICAEN) for critical
readings of the paper, and F. Ladieu (SPEC-CEA-SACLAY) for very
helpful remarks concerning the slow dynamic of magnetic systems.
H.W.E. acknowledges the CNRS and the conseil r$\acute{e}$gional de
Basse Normandie for funding.


\newpage

\begin{figure}[tbp]
\centering \includegraphics*[width=6cm]{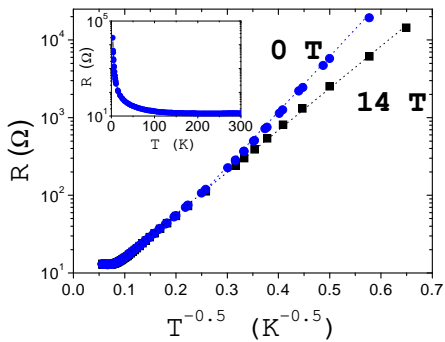} \caption{Color
online. Semilog plot of the resistance as function of $1/\sqrt T$,
for $B=0$ $T$ and $14$ $T$. Linearity in this scale relies on a
variable range hopping regime with a Coulomb gap. Inset:
Resistance as function of the temperature in a semi-log scale for
B=0T.} \label{fig.1}
\end{figure}

\begin{figure}[tbp]
\centering \includegraphics*[width=6cm]{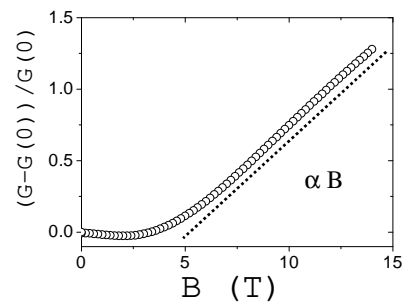}
\caption{Normalized magnetoconductance at $T=4.2$ $K$. Note the
small negative magnetoconductance for $B \lesssim 3$ $T$ followed
by the strong positive magnetoconductance. The dotted line is a
guide to the eyes.} \label{fig.2}
\end{figure}

\begin{figure}[tbp]
\centering \includegraphics*[width=8cm]{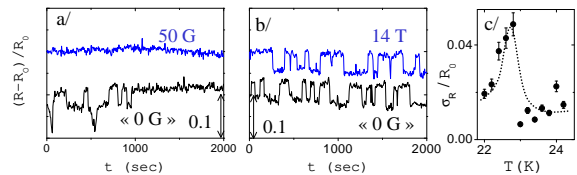} \caption{Color
online. a/ and b/ resistance traces at respectively 25 K and 22.8
K for $B=0$ and $5.10^{-3}$ T and for $B=0$ and $14$ $T$. The
shift of the time traces in each graph is arbitrary. $R_0$ is the
mean value of the resistance averaged during the time of the
measurement. c/ The normalized noise versus the temperature. Note
the peak of noise at $T=22.9 \pm 0.1$ $K$. The dotted line is a
guide to the eyes.} \label{fig.3}
\end{figure}
\begin{figure}[tbp]
\centering \includegraphics*[width=6cm]{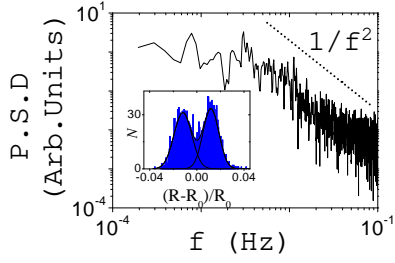}
\caption{Color online. The power spectal density (P.S.D)
corresponding to a very long recording of the time trace at
$T=22.80$ $K$. The well defined telegraph-like carriers dynamics
are evidenced by both the Lorentzian spectrum (corner frequency
$f_{c} \approx 10^{-2}$ Hz) and the bimodal histogram.}
\label{fig.4}
\end{figure}

\begin{figure}[tbp]
\centering \includegraphics*[width=6cm]{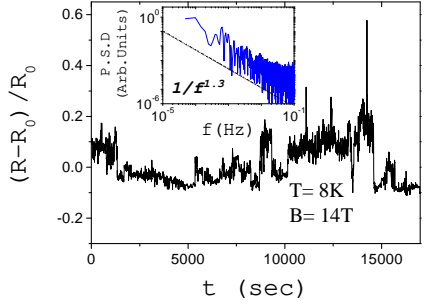} \caption{Color
online. Long time acquisition of the resistance trace at $T=8$ $K$
and $B=14$ $T$. Inset: The associated Fourier transform with a
($1/f^{1.3}$) form.} \label{fig.5}
\end{figure}

\begin{figure}[tbp]
\centering \includegraphics*[width=6cm]{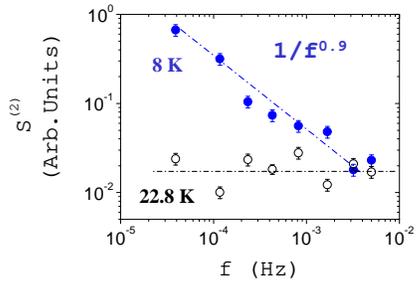} \caption{Color
online. Second spectra $S^{2}(f)$ of the resistance fluctuations
at 22.8 K and 8 K. Correlated fluctuations are evidenced by the
non-white second spectrum such as measured at the low
temperature.} \label{fig.6}
\end{figure}

\begin{figure}[tbp]
\centering \includegraphics*[width=6cm]{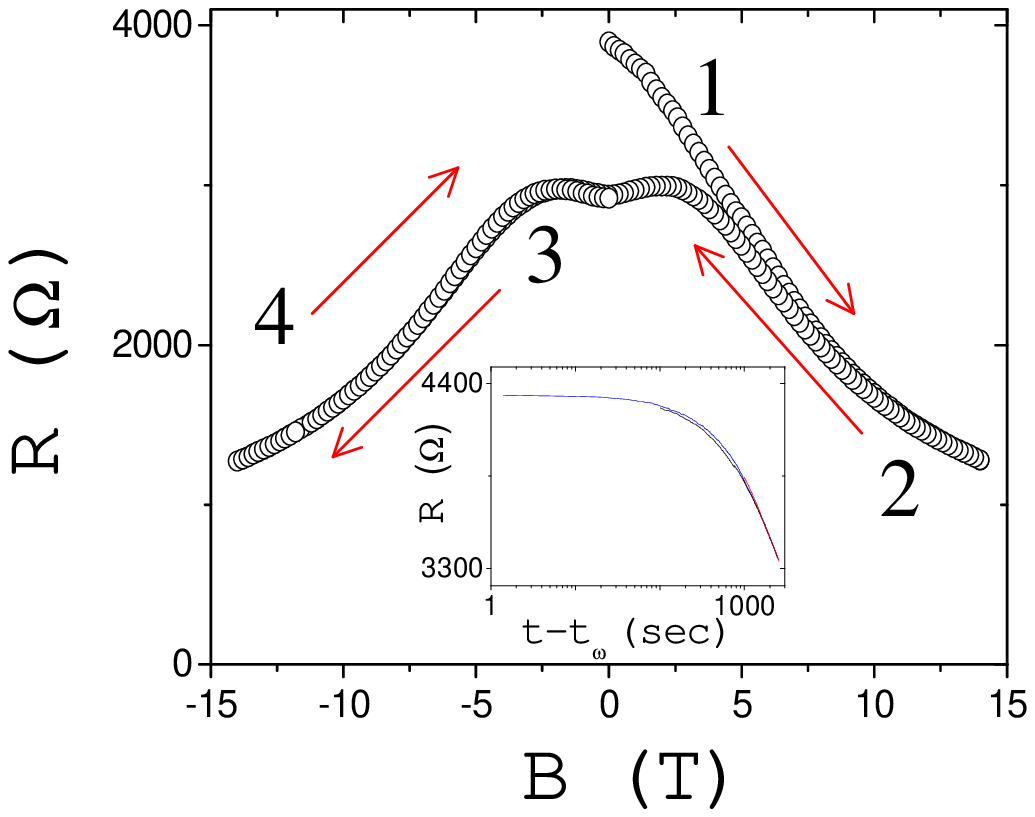} \caption{Color
online. Variation of the resistance as function of the magnetic
field at $T=4$ $K$ after a zero field cooling. Inset: relaxation
of the thermo-remanent resistance after a 2T FC for $t_{w}= 10,
100, 1000$ $sec$ as function of $t-t_w$.} \label{fig.7}
\end{figure}


\begin{references}
\label{sec:TeXbooks}

\bibitem{sylvie1} S. H$\acute{e}$bert, S. Lambert, D. Pelloquin, and A. Maignan , Phys. Rev. B 64, 172101
(2001).

\bibitem{supra}  K. Takada, H. Sakurai, E. Takayama-Muromachi, F. Izumi, R.A. Dilanian, and T. Sasaki, Nature (London) 422, 53 (2003).

\bibitem{ando} Y. Ando, N. Miyamoto, K. Segawa, T. Kawata, and I. Terasaki, Phys. Rev. B 60, 10580 (1999).

\bibitem{limelette} P. Limelette, V. Hardy, P. Auban-Senzier, D. J$\acute{e}$rome, D. Flahaut, S. H$\acute{e}$bert, R. Fr$\acute{e}$sard, Ch. Simon, J. Noudem, and A. Maignan, Phys. Rev. B 71, 233108 (2005).

\bibitem{riri} D. Grebille, S. Lambert, F. Bour$\acute{e}$e, and V. Petricek, J. Appl. Cryst. 37, 823 (2004).

\bibitem{koshi} W. Koshibae and S. Maekawa, Phys. Rev. Lett. 91, 257003
(2003).

\bibitem{sylvie} L. B. Wang, A. Maignan, D. Pelloquin, S. H$\acute{e}$bert, and B. Raveau , J. Appl. Phys. 92, 124 (2002).
 J. Sugiyama, C. Xia, and T. Tani, Phys. Rev. B 67, 104410 (2003).

\bibitem{Sujiyama} J. Sugiyama, J. H. Brewer, E. J. Ansaldo, H. Itahara, K. Dohmae, Y. Seno, C. Xia, and T. Tani, Phys. Rev. B 68, 134423 (2003).

\bibitem{mike} M. B. Weissman, Rev. Mod. Phys. 60, 537 (1988).

\bibitem{eng} H.W. Eng, W. Prellier, S. H$\acute{e}$bert, D. Grebille, L. M$\acute{e}$chin, and B. Mercey, J. Appl. Phys. 97, 013706
(2005).

\bibitem{masset} A. C. Masset, C. Michel, A. Maignan, M. Hervieu, O. Toulemonde, F. Studer, B. Raveau, and J. Hejtmanek, Phys. Rev. B 62, 166 (2000).

\bibitem{mott} N.F. Mott, Metal-Insulator Transitions (Taylor and
Francis, London), 1974.

\bibitem{zabro} A.G. Zabrodskii and K.N. Ninov'eva, Zh. Eksp. Teor. Fiz. 86, 727 (1984) [Sov. Phys. JETP 59, 425 (1984)].

\bibitem{Efros1} A.L. Efros and B.I. Shklovskii, J. Phys. C8, L49
(1975).

\bibitem{Efros2} B.I. Shklovskii and A.L. Efros, Electronic
Properties of doped semiconductor (Springer, New York, 1984).

\bibitem{nguyen} V.L. Nguyen, Sov. Phys. Semicond. 18, 207 (1984).

\bibitem{perez} A. P$\acute{e}$rez–Garrido, M. Ortu$\tilde{n}$o, E. Cuevas, J. Ruiz , and M. Pollak, Phys Rev B 55, R8630 (1997).

\bibitem{ra} M. E. Raikh and I. M. Ruzin, Sov. Phys. JETP 68, 642 (1989).



\bibitem{raquet} B. Raquet, M. N. Baibich, J. M. Broto, H. Rakoto, S. Lambert, and A. Maignan, Phys. Rev. B 65, 104442 (2002).


\bibitem{gruner} G. Gr$\ddot{u}$ner, Rev. Mod. Phys. 66, 1 (1994).


\bibitem{sivan} U. Sivan, O. Entin–Wohlman, and Y. Imry, Phys. Rev. Lett. 60, 1566 (1988).
O. Entin–Wohlman, Y. Imry, and U. Sivan, Phys. Rev. B 40, 8342
(1989).


\bibitem{nguyen2} V. I. Nguyen, B. Z. Spivak and B. I. Shklovskii, Zh. Eksp. Teor. Fiz. 89, 1770 (1985) [Sov. Phys—JETP 62, 1021 (1985)]

\bibitem{massey} J.G. Massey and M. Lee, Phys. Rev. Lett. 79, 3986
(1997).

\bibitem{restle} P. J. Restle, R. J. Hamilton, M. B. Weissman, and M. S. Love, Phys. Rev. B 31, 2254 (1985).

\bibitem{percol} G. T. Seidler, S. A. Solin, and A. C. Marley, Phys. Rev. Lett. 76, 3049
(1996).

\bibitem{bogda} S. Bogdanovich and D. Popovi$\acute{c}$, Phys. Rev. Lett. 88, 236401
(2002). J. Jaroszy$\acute{n}$ski, D. Popovi$\acute{c}$, and T. M.
Klapwijk, Phys. Rev. Lett. 89, 276401 (2002).

\bibitem{ladieu} F. Ladieu, F. Bert, V. Dupuis, E. Vincent, J. Hammann, J. Phys. Condens. Matter
16, 735 (2004).

\bibitem{vincent} E. Vincent, V. Dupuis, M. Alba, J. Hammann, and J.P. Bouchaud, Europhys. Lett. 50, 674 (2000).

\bibitem{sasaki}M. Sasaki, P.E. J$\ddot{o}$nsson, H. Takayama, and H. Mamiya, Phys. Rev. B 71, 104405 (2005).


\end{references}
\end{document}